\documentclass[journal]{IEEEtran}
\usepackage{cite}
\usepackage[pdftex]{graphicx}
\graphicspath{{Images/}}
\DeclareGraphicsExtensions{.pdf,.jpeg,.png,.eps}
\usepackage{epstopdf}
\usepackage{amsmath}
\usepackage{algorithmic}
\usepackage{array}
\usepackage{graphics}
\usepackage{adjustbox}
\usepackage{fixltx2e}
\usepackage{stfloats}
\usepackage{url}

\usepackage{booktabs}
\usepackage{longtable}
\usepackage{multicol}
\usepackage{caption}
\usepackage{multirow}
\usepackage{rotating}
\usepackage{float}
\usepackage{flushend}
\usepackage[referable]{threeparttablex}
\usepackage{xtab}
\usepackage{ragged2e}
\usepackage{fancyhdr}
\fancypagestyle{plain}{
  \fancyhf{}
  \fancyhead[L]{ASIAN JOURNAL OF ENGINEERING, SCIENCES \& TECHNOLOGY, VOL.5, ISSUE 1 \hfill MARCH 2015}     
    \fancyfoot[C]{\thepage}

}
\makeatletter
\def\ps@IEEEtitlepagestyle{%
  \def\@oddhead{\mycopyrightnotice}%
  \def\@evenhead{}%
}
\def\mycopyrightnotice{%
  {ASIAN JOURNAL OF ENGINEERING, SCIENCES \& TECHNOLOGY, VOL.5, ISSUE 1 \hfill MARCH 2015}
  \gdef\mycopyrightnotice{}
}
\begin{document}
\title{Brain Interface Based Wheel Chair Control System for Handicap \\ {\large An advance and viable approach}}
\author{Mohtashim Baqar\IEEEauthorrefmark{1}, Azfar Ghani\IEEEauthorrefmark{1}, Azeem Aftab\IEEEauthorrefmark{1} and Shahzad Karim Khawar\IEEEauthorrefmark{1}\thanks{Mohtashim Baqar, Azfar Ghani, Azeem Aftab and Shahzad Karim Khawar are with Faculty of Engineering, Sciences and Technology, IQRA University (Main Campus), Karachi (e-mail: mohtashim@iqra.edu.pk, azfar.ghani@iqra.edu.pk, azeem.cheema@iqra.edu.pk and shahzad.karim@iqra.edu.pk).}}
\maketitle
\pagestyle{plain}
\begin{abstract}
This paper presents advancement towards making an efficient and viable wheel chair control system based on brain computer interface via electro-oculogram (EOG) signals. The system utilizes the movement of eye as the element of purpose for controlling the movement of the wheel chair. Skin-surface electrodes are placed over skin for the purpose of acquiring the electro-oculogram signal and with the help of differential amplifier the bio-potential is measured between the reference and the point of interest, afterwards these obtained low voltage pulses are amplified, then passed through a sallen-key filter for noise removal and smoothening. These pulses are then collected on to the micro-controller; based on these pulses motor is switched to move in either right or left direction. A prototype system was developed and tested. The system showed promising results. The test conducted showed 99.5\% efficiency of movement in correct direction.
\end{abstract}
\IEEEpeerreviewmaketitle
\section{Introduction}
Bio-electricity being the fundamental essence of human life processes can be used to develop modern means of technology, working for the benefit of mankind. Human body possesses varying electrical potential on different nerves of the human body. These potential can be used to make predictions about health, fitness and can be used to treat physiological, psychological diseases and disorders.Recent advancements in computing and signal processing techniques have made it possible to develop a mind-machine interface (MMI). The use of EEG (electroencephalogram), EMG (electro-myogram), EOG (electro-oculogram), ECG (electro-cardiogram) or brain waves signals generated in human body are used to communicate with a computer \cite{dhillon2009eog} \cite{estrany2009eog}. These electrical signals are generated as a result of a thought to prompt a change or to prompt a muscle to move in a human body. The pulses generated propagate to different parts of the body through the nerve cells to prompt a change, acknowledging the thought that caused it. \par 
Stroke and spinal cord injuries are very ineffectual for human body \cite{bulling2008robust}; about fifteen percent of world population is affected directly or indirectly. It may reduce the productivity and performance of intellectual. Brain computer interface is one of the solutions which may tends to recover the productivity of people but, as it deals with the complicated human brain structure, required very extensive level of accuracy in signal processing of neural-signal \cite{harun2009eog,riccio2013attention,mattiocco2006neuroelectrical}. Another aspect of is to facilitate the handicap and old age person and provide continuous biomedical parameter monitoring.  Brain interface systems are used in many of the military and medical applications, made to work for the benefit of civil society. Table \ref{table:VR} give below, illustrates the bio-potential and frequency ranges of different types of signals, acquired from different parts of the human body.
\begin{table}[H]
\caption[Voltage Ranges of Bio-electric Signals]{Voltage Ranges of Bio-electric Signals}
\label{table:VR}
\centering
\begin{tabular}{c c c }
\toprule
\multicolumn{1}{ c }{\footnotesize{\bfseries Signals}}& \multicolumn{1}{ c }{\footnotesize{\bfseries Voltage Range}} & \multicolumn{1}{ c }{\footnotesize{\bfseries Frequency Range}}\\
\cmidrule{1-3}
Electroencephalogram & \multicolumn{1}{ c }{\footnotesize \bfseries 10 $\mu$V to 100$\mu$V} & \multicolumn{1}{ c }{\footnotesize \bfseries 0.1 Hz to 50 Hz}\\
Electro-oculogram & \multicolumn{1}{ c}{\footnotesize\bfseries 50$\mu$V to 3500$\mu$V} & \multicolumn{1}{ c }{\footnotesize \bfseries 1 Hz to 50 Hz} \\
Electro-cardiogram & \multicolumn{1}{c}{\footnotesize\bfseries 1 mV to 100 mV} & \multicolumn{1}{ c }{\footnotesize \bfseries 0.05 Hz to 100 Hz} \\
\bottomrule
\end{tabular}
\end{table}
The proposed system is designed to work using EOG signals, as part of the Human-Computer Interface (HCI). EOG signals are the ones originating in the eye. In recent years, the electro-oculography (EOG) has been set up as a promising methodology among different sources like electroencephalography (EEG) or electro-myography (EMG) \cite{fatourechi2007emg,deng2010eog}. Considering these factors various eye based applications have been suggested, acting as hands-free interfaces to a general PC framework or even wearable embedded system, with specific end goal to support the context awareness of machines.
\section{Working Principle}
The scope of the project involves the collection of EOG data, which is used for controlling the movement of the wheelchair. The procedure adapted in the proposed model is that electrodes are mounted on the surface of the skin, one being placed at reference point and other one placed at nerve fibre hold outside retina,the two points are fed into a differential amplifier, the acquired low energy electrical pulse is then amplified. The amplified signal is then fed into sallen-key low pass filter for noise removal and smoothing of the electrical pulses. The modified electrical pulses are then fed into the micro-controller. Serial communication takes place between the micro-controller and the system modules. The micro-controller continuously receives data from the low-polarization surface electrodes,and then takes decision on the premise of the readings collected.  A DC motor is connected to the micro-controller. The motor is witched in either of the two directions based on the readings collected on to the micro-controller. 
\begin{figure}[H]
\includegraphics[width=\columnwidth]{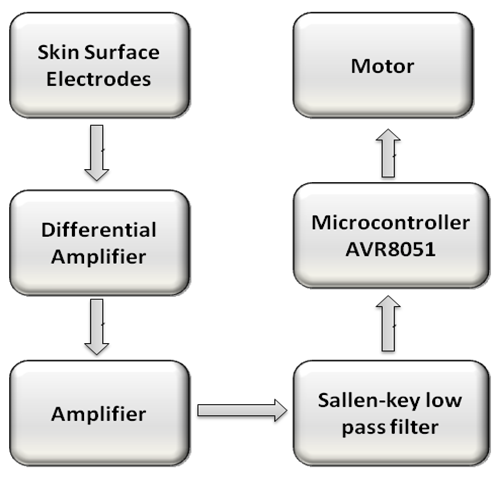}
\caption{Block Diagram of the proposed model}
\label{figure:BD}
\end{figure}
\subsection{Electrode}
Skin-surface electrodes made of Silver (Ag) are used for acquiring the EOG signals from the skin. The advantage of using silver is that when it comes in contact with a base it results into a slightly dis-solvable salt, i.e. silver chloride, which rapidly saturates and comes to equilibrium \cite{venkataramanan2005biomedical}. A cup-shaped electrode gives enough volume to carry an electrolyte, inclusive of chlorine ions. \par 
Generally electrodes do not make a conducting connection with the skin. A conducting gel is needed which is rubbed onto the skin before placing the electrodes, just to build-up the conducting process.
\subsection{Micro-controller}
Micro-controller 8052 is a very important part of this system. It acts like a computer on a chip. Its basic features are; \par
It has a ROM of 8 KB, RAM of 256 B, and three built-in timers. It is an economical and low power utilizing device. It can be directly programmed by connecting it with PC via serial link or program can be written offline on a PC then can be uploaded on a micro-controller. \par                 
Moreover, on acquisition of EOG signal, the controller prompts motor to switch in either left or right direction.
\subsection{Differential Amplifier}
The utilization of differential amplifier is common in bio-potential estimations in light of a more noteworthy capacity to dismiss ecological obstruction compared with ground referenced single ended amplifiers \cite{wang2005design}. \par 
Differential amplifier deducts the electric potential available at one place on the body from the other one. A third body location is set as a reference point that serves as a common point; both potentials are measured with respect to it. Differential amplifiers becomes more useful here because the bio-potentials created inside the body changes over the body surface, however line-coupled noise does not. Further, output of the differential amplifier can be calculated as described in equation \ref{eq:DA}. \par
\begin{equation}
v_{out}=A(v_p-v_{ref}) 
\label{eq:DA}
\end{equation}
where, $V_{out}$ is the output voltage of the differential amplifier, is the gain of the amplifier, $V_p$ voltage at point of interest and $V_{ref}$ is the voltage at the reference point of the body.\par 
Further, the acquired EOG signals are fed into a differential amplifier and the difference of it is fed to an amplifier.
\subsection{Amplifier}
The Burr-brown Ina114 instrumentation amplifier is used in this project. This chip has high input impedance that functions better with even high impedance bio-electrodes. The amplifier gain can be chosen from several standard gains by interconnection of pins on the package or alternatively, by changing the value of a single external resistor \cite{lopeandia2008power}. \par 
The amplifier is used to amplify the low energy bio-potential signal acquired via skin surface electrodes.
\subsection{Sallen-Key Filters}
Sallen-key is a 2nd order filter. It is a degenerated form voltage controlled voltage oscillator. It is a second order filter. The Sallen-key filter gives a distinguished feature in providing high gain and it is less prone to noise \cite{saraga1967sensitivity}. It is one of the few models which can provide large gain while using minimum number of operational amplifiers. \par 
In this proposed model Sallen key low pass filters are used. It's been used for the purpose of high frequency noise components from the desired output signal which is fed into the micro-controller.
\subsection{Motor}
A 12V DC motor is used, which is attached to the wheels of the wheelchair. The EOG data collected is fed to the micro-controller which prompts the motor to move either in right or left direction. \par 
Supporting hardware or circuitry like power supply and snubber circuit is used. For controlling and limiting the surge of current which passes through various modules like operational amplifier and differential amplifier, Snubber circuits are used. The power requirement of the system will be fulfilled by drawn it from a portable battery that is placed under a wheelchair.To limit the current as per the requirement of the system, a rheostat is deployed in between the battery and the system.
\section{programming interface}
Assembly language is being used for micro-controller programming. The micro-controller works on the basis of the EOG signal received via skin-surface electrodes.
\subsection{Assembly Language} 
To program micro-controller using assembly language is one of the best options because it is a low level programming language, therefore its each code is related to exactly one of micro-controller's machine level instructions. It utilizes mnemonic codes to locate data on physical memory instead of remembering exact location of data.
\subsection{SYSTEM IMPLEMENTATION AND RESULTS}
Complete circuitry with all hardware was integrated on a wheel chair. The system was tested with 5 different subjects with different eye sight. Around 1000 samples were collected for each person. Observed results for each individual showed little variation in terms of accuracy of correct movement. Overall results gave an average accuracy of almost 98.90\%. Further, table \ref{table:CM} shows an average percentage accuracy of correct movement along with an estimated value of variance for each individual subject.
\begin{table}[H]
\caption[Voltage Ranges of Bio-electric Signals]{Voltage Ranges of Bio-electric Signals}
\label{table:CM}
\centering
\begin{tabular}[width=\columnwidth]{p{2in} p{2in} p{2in} }
\toprule
\multicolumn{1}{ c }{\footnotesize{\bfseries Subjects}}& \multicolumn{1}{ c }{\footnotesize{\bfseries Accuracy}} & \multicolumn{1}{ c }{\footnotesize{\bfseries Variance}}\\
\cmidrule{1-3}
\multicolumn{1}{ c }{\footnotesize \bfseries 01}& \multicolumn{1}{ c }{\footnotesize \bfseries $98.00\%$} & \multicolumn{1}{ c }{\footnotesize \bfseries $\pm 1\%$}\\
\multicolumn{1}{ c }{\footnotesize \bfseries 02}& \multicolumn{1}{ c}{\footnotesize\bfseries $98.70\%$} & \multicolumn{1}{ c }{\footnotesize \bfseries $\pm 1.2\%$}\\
\multicolumn{1}{ c }{\footnotesize \bfseries 03}& \multicolumn{1}{c}{\footnotesize\bfseries $99.50\%$} & \multicolumn{1}{ c }{\footnotesize \bfseries $\pm 1\%$}\\
\multicolumn{1}{ c }{\footnotesize \bfseries 04}& \multicolumn{1}{ c}{\footnotesize\bfseries $99.00\%$} & \multicolumn{1}{ c }{\footnotesize \bfseries $\pm 0.5\%$}\\
\multicolumn{1}{ c }{\footnotesize \bfseries 05}& \multicolumn{1}{c}{\footnotesize\bfseries $98.30\%$} & \multicolumn{1}{ c }{\footnotesize \bfseries $\pm 1.5\%$}\\
\bottomrule
\end{tabular}
\end{table}
\section{Acknowledgement}
The authors would like to acknowledge the Faculty of Engineering, Sciences and Technology (FEST), IQRA University for their persistent support and encouragement throughout the course of the project.
\section{Conclusion}
An advance wheelchair control system has been proposed and a prototype has been built based on EOG signals. Electrodes are mounted to acquire EOG data. The movement of the wheelchair is controlled making use of potential difference between optic fibre nerve and the reference point. The low voltage pulse is amplified and fed into the micro-controller on the basis of the decision is taken, whether to switch the motor in right or in the left direction. As the person moves his eyes in the right direction the wheelchair moves towards right and vice versa. In future this system can be further enhanced with the use of better sensors/electrodes and can be made to produce much more efficient results.  
\bibliographystyle{IEEEtran}
\bibliography{ref}
\vspace{-0.4in}
\begin{IEEEbiographynophoto}{Mohtashim Baqar}
received the B.E and M.S degrees in Telecommunication and Electrical Engineering from IQRA University (Main Campus), Karachi and National University of Sciences and Technology, Islamabad, Pakistan, in 2011 and 2016, respectively. Since 2012, he has been a Lecturer in Faculty of Engineering, Sciences and Technology, IQRA University (Main Campus), Karachi. His research interests are multi-variate signal processing and pattern recognition.
\end{IEEEbiographynophoto}
\vspace{-0.4in}
\begin{IEEEbiographynophoto}{Azfar Ghani}
received the B.E degree in Electronic Engineering from IQRA University (Main Campus), Karachi 2012. Since 2011, he has been a Lab Engineer in Faculty of Engineering, Sciences and Technology, IQRA University (Main Campus), Karachi. His research interests are power electronics and industrial electronics. 
\end{IEEEbiographynophoto}
\vspace{-0.4in}
\begin{IEEEbiographynophoto}{Azeem Aftab}
received the B.E and M.S degrees in Electronic Engineering and Telecommunication from IQRA University (Main Campus), Karachi in 2010 and 2016, respectively. He was with Etilize, Pakistan from 2011 to 2012. Since 2013, he has been a Lecturer in Faculty of Engineering, Sciences and Technology, IQRA University (Main Campus), Karachi. His research interests are renewable energy and radiating systems.  
\end{IEEEbiographynophoto}
\vspace{-0.4in}
\begin{IEEEbiographynophoto}{Shahzad Karim Khawar}
received the B.S and M.S degrees in Computer Sciences from Usman Institute of Technology, Hamdard University, Karachi, in 2006 and 2010, respectively. He was with Future Tech Services, Pakistan from 2007 to 2012. Since 2013, he is with the Faculty of Engineering, Sciences and Technology, IQRA University (Main Campus), Karachi. His research interests are Computer Networks and Dynamic Programming Languages.  
\end{IEEEbiographynophoto}
\end{document}